\documentclass[12pt]{article}

\usepackage[margin=1in]{geometry}
\usepackage{amsmath,amssymb,amsfonts}
\usepackage{hyperref}

\usepackage{times}
\usepackage[T1]{fontenc}
\usepackage{graphics, graphicx}
\usepackage{color}
\usepackage{subfigure}
\usepackage{textpos}
\usepackage{float}
\usepackage{authblk}
\usepackage[english]{babel}
\usepackage{caption}
\usepackage[normalem]{ulem}

\DeclareGraphicsExtensions{.pdf,.png,.jpg}

\allowdisplaybreaks

\title{Estimating Seroprevalence of SARS-CoV-2 in Ohio: A Bayesian Multilevel Poststratification Approach with Multiple Diagnostic Tests}

\date{}

\author[1]{David Kline}
\author[2]{Zehang  Li}
\author[3]{Yue Chu}
\author[4]{Jon Wakefield}
\author[5]{William C. Miller}
\author[6]{Abigail Norris Turner}
\author[3]{Samuel J. Clark}

\affil[1]{Center for Biostatistics, Department of Biomedical Informatics, The Ohio State University, Columbus, Ohio}
\affil[2]{Department of Statistics, University of California Santa Cruz, Santa Cruz, California}
\affil[3]{Department of Sociology, The Ohio State University, Columbus, Ohio}
\affil[4]{Department of Statistics and Department of Biostatistics, University of Washington, Seattle, Washington}
\affil[5]{Division of Epidemiology, College of Public Health, The Ohio State University, Columbus, Ohio}
\affil[6]{Division of Infectious Diseases, College of Medicine, The Ohio State University, Columbus, Ohio}

\begin{document}

\maketitle

\begin{abstract}
Globally the SARS-CoV-2 coronavirus has infected more than 59 million people and killed more than 1.39 million. Designing and monitoring interventions to slow and stop the spread of the virus require knowledge of how many people have been and are currently infected, where they live, and how they interact. The first step is an accurate assessment of the population prevalence of past infections. There are very few population-representative prevalence studies of the SARS-CoV-2 coronavirus, and only two American states -- Indiana and Connecticut -- have reported probability-based sample surveys that characterize state-wide prevalence of the SARS-CoV-2 coronavirus. One of the difficulties is the fact that the tests to detect and characterize SARS-CoV-2 coronavirus antibodies are new, not well characterized, and generally function poorly. During July, 2020, a  survey representing all adults in the State of Ohio in the United States collected biomarkers and information on protective behavior related to the SARS-CoV-2 coronavirus. Several features of the survey make it difficult to estimate past prevalence: 1) a low response rate, 2) very low number of positive cases, and 3) the fact that multiple, poor quality serological tests were used to detect SARS-CoV-2 antibodies. We describe a new Bayesian approach for analyzing the biomarker data that simultaneously addresses these challenges and characterizes the potential effect of selective response. The model does not require survey sample weights, accounts for multiple, imperfect antibody test results, and characterizes uncertainty related to the sample survey and the multiple, imperfect, potentially correlated tests.
\end{abstract}

\section{Introduction} 
\label{sec:intro}

Slowing or stopping the spread of a new virus for which a vaccine does not exist starts with two key pieces of information.  One, what fraction of the population has been infected and is thereby potentially less or not susceptible to future infection, and two, what fraction of the population is currently infected and potentially able to infect others. Together with a basic understanding of the infection process, this information roughly characterizes the potential for the epidemic to grow.  This is invaluable to public health officials and policy makers who have the responsibility to manage the epidemic and protect the public.

As of this writing (late November, 2020), the global SARS-CoV-2 coronavirus (hereafter ``CV19'') pandemic has infected more than 59 million people and killed more than 1.39 million \cite{dong2020interactive}. Basic epidemiological information to describe the pandemic is scarce because the virus is new and the pandemic exploded rapidly. In its place are a wide variety of indicators based on convenient, mostly non-representative, or indirectly related data -- counts of all-cause deaths (e.g. \cite{national2020excess,weinberger2020estimation,rossen2020excess}), facility-based testing results for symptomatic patients, non-representative samples, and in many situations, results from inadequately characterized tests that perform poorly \cite{bastos2020diagnostic}. Franceschi and colleagues \cite{franceschi2020population} identify 37 CV19 prevalence studies from nineteen countries.  Most present results from non-representative, otherwise special, or very small study populations.  Just fourteen represent large enough populations to be of policy interest -- national or state-level -- and use a probability-based sample from a credible sampling frame to produce results that could represent the population of interest -- Asia: one \cite{shakiba2020seroprevalence}, Europe: seven \cite{ward2020antibody,riley2020community,vodivcar2020low,merkely2020novel,pollan2020prevalence,snoeck2020prevalence,gudbjartsson2020spread}, North America: two \cite{menachemi2020population,mahajan2020seroprevalence}, and South America: four \cite{gomes2020population,da2020population,silveira2020population,hallal2020remarkable}. The two in North America are the State of Indiana \cite{menachemi2020population} and the State of Connecticut \cite{mahajan2020seroprevalence} in the United States.  

Conducting population-representative biomarker surveys is difficult -- particularly in the United States. Good sampling frames exist in a variety of forms (tax rolls, telephone numbers, etc.), but recruiting willing respondents is exceptionally difficult and likely affected by selection relative to the outcome of interest. Both of the studies in the United States had low response rates for the full interview with valid CV19 test results -- Indiana: 23.4\% and Connecticut: 7.8\%. Further complicating analysis, there were few positive tests among those who did respond -- Indiana: $47/3,658 = 1.3\%$ for the PCR test of current infection, and $38/3,658=1.0\%$ for antibody test of ever infected; and Connecticut: $23/567 = 4.1\%$ for the antibody test of ever infected. Both studies described concern that the non-responding participants were likely to be at higher risk of infection with CV19.  Finally, like all CV19 immunology investigations to date, both studies struggled with poor quality antibody tests whose unfavorable performance characteristics were not well understood; see \cite{shakiba2020seroprevalence} for an overview of these issues.  

Statistical analysis of data like these is difficult. First, the low response rate requires extensive recalibration of the sampling weights, and in the worst case, there may be sampling units with no respondents at all. Second, the very small number of positive cases pushes the asymptotic (large sample) assumptions of frequentist methods to their limits and can break them. Third, the imperfect and poorly characterized antibody tests potentially add a lot of uncertainty that must be reflected in the results, particularly in low prevalence settings \cite{Gelman2020}. Fourth, when results from multiple tests with different performance characteristics are combined, the joint result must be accurately described and its uncertainty propagated to the final estimate of prevalence -- importantly, including the possibility that results from individual tests are correlated. Finally, if there is selection on the outcome, then the effect of this must be understood. In our review of the literature we did not find an existing method that addresses all of these challenges in a unified way.

Here, we describe an analytical approach developed to produce estimates of past infection with CV19 using data from a probability-based household survey representing adults in the State of Ohio in the United States. Like the CV19 prevalence studies in Indiana and Connecticut, the Ohio survey had a low response rate, few positive cases, and the possibility of selective response. Additionally, the Ohio survey used multiple imperfect antibody tests for the same antibodies, resulting in the need to quantify uncertainty in the joint result and account for possible dependence among results.  


To overcome these challenges, we weave together two well-established modeling frameworks into a single coherent approach.  We utilize the literature on modeling multiple imperfect diagnostic tests through the use of a Bayesian latent class model (e.g. \cite{Dendukuri2001,Wang2017}).  This enables us to combine information across tests to infer the true latent infection status of a participant while incorporating uncertainty about the characteristics of the tests.  We use the latent infection status to generate model-based estimates of the population prevalence using multilevel regression and poststratification \cite{Gelman2007,Gelman2020}. These approaches are integrated into a single Bayesian model that allows for the full propagation of uncertainty, exact inferences, and the ability to specify informative priors using external information.  By doing so we produce estimates that reflect all available information and uncertainty.

\section{Methods}
\label{sec:methods}

The purpose of this study is to estimate the prevalence of past CV19 infections in the State of Ohio using three separate antibody tests given to randomly selected adult participants.  We know that each antibody test is imperfect, and there is no gold standard for detecting prior CV19 infection.  Prevalence estimates based on a single imperfect test are always biased, but particularly in the case of CV19 infection rates, which are low \cite{Gelman2020}.  To mitigate that bias and incorporate variability due to error in the testing results, we will take a Bayesian latent class approach for modeling multiple diagnostic tests.  Our approach will be based on combining a fixed effects framework for modeling conditional dependence across multiple diagnostic tests \cite{Dendukuri2001,Wang2017} with a model-based analysis using multilevel regression and poststratification \cite{Gelman2007} to acknowledge the complex design aspects of the survey.

\subsection{Survey Design}

The survey was designed to provide policymakers with a ``quick'' overall snapshot of the prevalence of prior infection at the state level.  The survey sampling scheme was designed as a stratified two-stage cluster sample.  Strata were defined by 8 administrative regions used by the state.  Within each region, 30 census tracts were randomly selected with probability proportional to size (PPS) based on total population size.  Then within a selected tract, 5 households were randomly selected and one adult (at least 18 years of age) within each household was randomly selected to participate in the study.  Thus, the planned target sample size was 1200 participants.  The study was conducted from July 9-28, 2020.

Each participant in the study provided biological samples which would be put through a series of diagnostic tests. In this paper, we will focus on prior infection as determined by the presence of SARS-CoV-2 antibodies.  Since antibody tests for CV19 are new to the market and of varying quality, we processed participant samples using 3 different antibody tests.  Specifically, we used the Abbott IgG, Liaison IgG, and Epitope IgM tests.

\subsection{Model}

For each participant in the study, indexed by $i=1,\ldots,n$, let $\textbf{T}_i=(T_{i1}, T_{i2}, T_{i3})$ be indicators of a positive test result of the Abbott IgG, Liaison IgG, and Epitope IgM, respectively. Let $D_{i}$ be the unobserved indicator of whether participant $i$ had a prior infection of CV19. This latent indicator of prior infection is our primary outcome of interest.  Analysis methods for multiple diagnostic tests without a gold standard hinge on assumptions related to conditional independence \cite{Wang2017}.  We will assume that $(T_{i1}, T_{i2})$ and $T_{i3}$ are independent given the true infection status.  This implies that conditional on the true presence of prior infection, we assume tests for the same antibody are dependent and tests for different antibodies are independent.  Based on the underlying design of the tests and what they target, we believe these assumptions are reasonable for this specific application.

\begin{table}
    \centering
    \caption{Mapping from the three binary test results to the multinomial response vector and corresponding probabilities.}
    \begin{tabular}{ccccc}
        $Y_i$ & $T_{i1}$ & $T_{i2}$ & $T_{i3}$ & $\textbf{p}_i$ \\
        \hline
        1 & 1 & 1 & 1 & $p_{i1}$\\
        2 & 1 & 1 & 0 & $p_{i2}$\\
        3 & 1 & 0 & 1 & $p_{i3}$\\
        4 & 0 & 1 & 1 & $p_{i4}$\\
        5 & 1 & 0 & 0 & $p_{i5}$\\
        6 & 0 & 1 & 0 & $p_{i6}$\\
        7 & 0 & 0 & 1 & $p_{i7}$\\
        8 & 0 & 0 & 0 & $p_{i8}$\\
        \hline
    \end{tabular}
    \label{tab:y}
\end{table}

Given the assumptions stated above, we can consider this a problem with two conditionally independent sets of tests: the two IgG tests and the IgM test.  Thus, we can decompose the joint probability as
\begin{align*}
    \Pr(T_{i1},T_{i2},T_{i3}|D_{i})=\Pr(T_{i1},T_{i2}|D_{i})\Pr(T_{i3}|D_{i}).
\end{align*}
By doing so, each conditional probability on the right hand side can be estimated following a fixed effects approach \cite{Dendukuri2001}. Since each test result is binary, this leads to 8 potential combinations of results, shown in Table \ref{tab:y}, and suggests the following distribution:
\begin{align*}
    Y_i | D_{i} \sim \text{Multinomial}(1, \textbf{p}_i)
\end{align*}
where $Y_i$ is an indicator of participant $i$'s result pattern and $\textbf{p}_i=\textbf{p}_i(D_i)$ is a vector of length 8 where each element is the probability of a result pattern.

To construct the probability vector $\textbf{p}_i$, we first need to define the conditional probabilities within each antibody.  Let $S_j$ be the sensitivity and $C_j$ be the specificity of test $j$.  Since we have two IgG tests ($j=1,2$), we allow for their results to be correlated. Let $R^1_{12}$ be the covariance between the results of test $1$ and $2$ given the infection status is positive and $R^0_{12}$ be the covariance when the infection status is negative.  Then for the joint probabilities of the IgG test results \cite{Dendukuri2001}, we have
\begin{align}
    \label{eq:twotests}
    \Pr(T_{i1}=t_{i1},T_{i2}=t_{i2}|D_{i}=1) = S_1^{t_{i1}} S_2^{t_{i2}} (1-S_1)^{1-t_{i1}} (1-S_2)^{1-t_{i2}} + (-1)^{t_{i1}+t_{i2}} R^1_{12} \\
    \Pr(T_{i1}=t_{i1},T_{i2}=t_{i2}|D_{i}=0) = C_1^{1-t_{i1}} C_2^{1-t_{i2}} (1-C_1)^{t_{i1}} (1-C_2)^{t_{i2}} + (-1)^{t_{i1}+t_{i2}} R^0_{12}. \nonumber
\end{align}
Since we only have one IgM test ($j=3$), we have
\begin{align}
    \label{eq:onetest}
    \Pr(T_{i3}=t_{i3}|D_{i}=d_{i})=d_{i}S_3^{t_{i3}}(1-S_3)^{1-t_{i3}}+(1-d_{i})C_3^{1-t_{i3}}(1-C_3)^{t_{i3}}.
\end{align}
We then use these conditional probabilities to construct the probabilities in the vector $\textbf{p}_i$ in the multinomial distribution above.  Those eight probabilities can calculated with the following general equation for each test result pattern
\begin{align*}
    \Pr(T_{i1}=t_{i1},T_{i2}=t_{i2},T_{i3}=t_{i3}|D_i=d_{i}) = \Pr(t_{i1},t_{i2}|d_{i}) \Pr(t_{i3}|d_{i}).
\end{align*}
These probabilities are the individual elements of the vector $\textbf{p}_i$ which is a function of $D_i,\textbf{S},\textbf{C},\textbf{R}$ where $\textbf{S}=(S_1,S_2,S_3)$, $\textbf{C}=(C_1,C_2,C_3)$ and $\textbf{R}=(R^1_{12},R^0_{12})$.

The latent true infection status is the primary process of interest as we are interested in estimating the prevalence of prior CV19 infections.  We assume
\begin{align*}
    D_{i}|\pi_{i} \sim \text{Bernoulli}(\pi_{i})
\end{align*}
where $\pi_{i}$ is the probability of prior infection for participant $i$, which will be determined by region, strata, and census tract. Specifically, we let $r[i]$, $s[i]$, and $t[i]$ refer to the region, strata, and census tract for participant $i$ and assume
\begin{align}
\label{eq:multilevel}
    \text{logit}(\pi_{i})=\alpha_{r[i]}+\textbf{X}_{s[i]}\boldsymbol{\beta}^s+\textbf{X}_{t[i]}\boldsymbol{\beta}^t+b_{t[i]}
\end{align}
where $\alpha_{r}$ is a region-specific random intercept, $\textbf{X}_{s}$ is a vector of stratum indicators with fixed effects vector $\boldsymbol{\beta}^s$, $\textbf{X}_{t}$ is a vector of census tract covariates with fixed effects vector $\boldsymbol{\beta}^t$, and $b_{t}$ is a random effect for census tract. In this analysis, $\textbf{X}_{s}$ includes indicators of participant age group and sex and $\textbf{X}_{t}$ is a scalar and corresponds to the log of the total census tract population. We code the groups using sum to 0 contrasts, and the log population was standardized to have 0 mean and standard deviation of 1 across the entire state. We assume $\alpha_{r} \stackrel{iid}{\sim} N(\alpha,\sigma^2)$.  Since the fixed effects are coded to sum to 0, $\alpha$ reflects the overall mean on the logit scale and $\alpha_r$ can be interpreted as regional means.  We also assume $b_{t} \stackrel{iid}{\sim} N(0,\tau^2)$, which accounts for correlation between individuals in the same census tract.

At this point, we link the diagnostic test model to a multilevel regression and poststratification approach \cite{Gelman2007,Gelman2020} for estimating population prevalence. In Equation [\ref{eq:multilevel}], we specified a multilevel logistic regression model for the probability of prior infection. Since true infection was rare in our sample, we were unable to fit a saturated model. Instead, we chose to use a hierarchical model with fixed strata effects. By doing so, we assume that while regions may have different probabilities of infection, the relative ordering of the strata will be the same across regions (i.e., there is no interaction between region and strata).   Effects for region and census tract population are included to account for the characteristics of the underlying survey design (stratification and PPS sampling), making the selection process ignorable \cite{Makela2018}.

To obtain the population prevalence, we calculate
\begin{align*}
    \pi=\frac{\sum_{r} \sum_{s} \sum_{t} \pi_{rst}P_{rst}}{\sum_{r} \sum_{s} \sum_{t}P_{rst}}
\end{align*}
where $\pi_{rst}$ is the prevalence and $P_{rst}$ is the adult population in stratum $s$ in Census tract $t$ in region $r$.  The prevalence contribution for region $r$, strata $s$ and tract $t$ is 
\begin{align*}
\pi_{rst} = \mbox{expit}( \alpha_{r}+\textbf{X}_{s}\boldsymbol{\beta}^s+\textbf{X}_{t}\boldsymbol{\beta}^t+b_{t})
\end{align*}
where $b_t \stackrel{iid}{\sim} N(0,\tau^2)$.

Since we are fitting the model in the Bayesian paradigm, we must specify prior distributions on all unknown parameters. The main reason we chose a fixed effects model for the test results was because it is directly determined by test sensitivity and specificity.  This allows us to transparently incorporate prior information using the validation data on the package insert (as of October 2020) for each test.  Based on this information, we let
\begin{align*}
    S_1 \sim \text{Beta}(109,13) \hspace{0.5in} C_1 \sim \text{Beta}(1066,4) \\
    S_2 \sim \text{Beta}(96,39) \hspace{0.5in} C_2 \sim \text{Beta}(1074,16) \\
    S_3 \sim \text{Beta}(9,11) \hspace{0.5in} C_3 \sim \text{Beta}(54,0.1).
\end{align*}
The remaining parameters for the diagnostic testing part of the model are the covariance parameters.  We enforce necessary constraints on these parameters by specifying independent uniform prior distributions for each parameter restricted to its allowable range \cite{Dendukuri2001}.  Assuming only positive dependence between tests, we have
\begin{align*}
    R^1_{12} \sim \text{Uniform}(0,\text{min}(S_1,S_2)-S_1 S_2)\\
    R^0_{12} \sim \text{Uniform}(0,\text{min}(C_1,C_2)-C_1 C_2).
\end{align*}
For the multilevel regression part of the model, we assume each element of $\boldsymbol{\beta}$ is independently normally distributed with 0 mean and variance of 9. We assume $\tau$ and $\sigma$ have independent uniform distributions on $(0,5)$. We assume $\alpha \sim N(\text{logit}(0.03),1)$ to reflect prior belief that the prevalence of prior infection is around 3\% and puts 95\% of the probability between 0.4\% and 18.0\%.

By putting everything together, we have the full model
\begin{align*}
p( \mathbf{S},\mathbf{C},\mathbf{R},\mathbf{D} ,\boldsymbol{\alpha},\boldsymbol{\beta},\mathbf{b},\tau^2 ,\alpha,\sigma^2| \mathbf{T})
&\propto \prod_{i=1}^n \Pr(\mathbf{T}_i |  \mathbf{S},\mathbf{C},\mathbf{R},D_i) \times p(\mathbf{R}|  \mathbf{S},\mathbf{C}) \times p(  \mathbf{S},\mathbf{C})\\
&\times \prod_{i=1}^n  \Pr ( D_i  | \boldsymbol{\alpha},\boldsymbol{\beta},\mathbf{b}) \times p(\mathbf{b} | \tau^2) \times p(\boldsymbol{\alpha} | \alpha,\sigma^2) \times p({\boldsymbol{\beta}}) p(\alpha) p(\tau^2) p(\sigma^2)
\end{align*}
where $\textbf{T}=(\textbf{T}_1,\ldots,\textbf{T}_n)$, $\textbf{S}=(S_1,S_2,S_3)$, $\textbf{C}=(C_1,C_2,C_3)$, $\textbf{R}=(R^1_{12},R^0_{12})$, $\textbf{D}=(D_1,\ldots,D_n)$, $\boldsymbol{\beta}=(\boldsymbol{\beta}^s,\boldsymbol{\beta}^t)$, $\boldsymbol{\alpha}$ is the vector of region-specific intercepts, and $\textbf{b}$ is the vector of census tract random effects. To compute the posterior distribution, the model was fit using a Markov Chain Monte Carlo (MCMC) algorithm implemented in R \cite{R} using NIMBLE \cite{nimble}.  The algorithm was run for 500,000 iterations, discarding the first 250,000 as burn-in, and thinning the remaining iterations by keeping every 20\textsuperscript{th} draw.  Convergence was assessed by visually inspecting trace plots.  Posterior distributions are summarized by the posterior mean and 95\% highest posterior density (HPD) credible interval. Code is available online at: \url{https://github.com/sinafala/bayes-prevalence}.

\subsection{Missing Test Results}

Some participants did not have results for all 3 tests considered. However, since our primary interest is in the latent infection status of the participant, we can still gain some information about this from the test results that are available.  Test result pattern probabilities would still be calculated as above corresponding to the test or pair of tests available for the participant.  For example, if a participant only had the IgG test results, there would be 4 patterns of results that would be defined by the probabilities in Equation [\ref{eq:twotests}].  If an IgG and IgM test are observed, the 4 probabilities would be computed from products of
\begin{align}
\label{eq:onetestgeneral}    
    \Pr(T_{ij}=t_{ij}|D_{i}=d_{i})=d_{i}S_j^{t_{ij}}(1-S_j)^{1-t_{ij}}+(1-d_{i})C_j^{1-t_{ij}}(1-C_j)^{t_{ij}}.
\end{align}
A single available test result would simply lead to a Bernoulli distribution of whether the result was positive ($t_{ij}=1$) with the probability calculated as in Equation [\ref{eq:onetestgeneral}] with the appropriate parameters for the observed test.

\subsection{Sensitivity Analyses}

For the analysis of any survey, it is important to account for the potential of non-response bias.  We note that our primary analysis is valid assuming a non-response mechanism that is ignorable given stratum.  This is likely a strong assumption and so we will consider sensitivity analyses under several assumptions of potential non-ignorable scenarios.  To do so, assume
\begin{align}
    \pi_{rst}=\pi^R_{rst} (1-p^N_{rst}) + \pi^N_{rst} p^N_{rst}
\end{align}
where $\pi^R_{rst}$ and $\pi^N_{rst}$ are the prevalence among the responders and non-responders and $p^N_{rst}$ is the probability of non-response.  Note that when ignorability is satisfied, we have $\pi_{rst}=\pi^R_{rst} = \pi^N_{rst}$, which results in the estimates from the primary analysis.  Due to the survey design, we can only obtain household non-response rates by region, and so, we assume $p^N_{rst}=p^N_{r}$ for all strata and tracts.  In order to carry out a sensitivity analysis, we will assume the prevalence among non-responders is: 
\begin{align}
\label{eq:non}
    \pi^N_{rst}=\lambda \pi^R_{rst}
\end{align}
where $\lambda$ is the prevalence ratio in non-responders compared to responders and $\pi^R_{rst}$ is the prevalence estimated from the logistic regression described above.  We will vary $\lambda$ to explore different scenarios and assess the sensitivity of our estimates to changes in the prevalence in non-responders.

\section{Results}
\label{sec:results}

\begin{table}
    \centering
    \caption{Descriptive statistics of participants with at least one antibody test result who were included in the analysis of prior infection prevalence.}
    \begin{tabular}{lrr}
        Variable & Count & Proportion \\
        \hline
        Age 18-44 & 175 & 0.262 \\
        Age 45-64 & 239 & 0.358 \\
        Age 65 and over & 253 & 0.379 \\
        \\
        Male & 275 & 0.412 \\
        Female & 392 & 0.588 \\
        \\
        Central & 82 & 0.123 \\
        East Central & 90 & 0.135 \\
        Northeast & 75 & 0.112 \\
        Northwest & 156 & 0.234 \\
        Southeast & 77 & 0.115 \\
        Southeast Central & 56 & 0.084 \\
        Southwest & 65 & 0.097 \\
        West Central & 66 & 0.099 \\
        \hline
    \end{tabular}
    \label{tab:sum_past}
\end{table}

A total of 727 adults participated in the survey. To be included in the analysis, participants had to have at least one antibody test result and have age and sex recorded.  Of the 727 participants, 667 (92\%) were included in the analysis. Characteristics of those included in the analysis are shown in Table \ref{tab:sum_past}. We observe that our included participants tend to be older and female. We also observe that 23.4\% of included participants were from the Northwest region of Ohio.

\begin{figure}
    \centering
    \subfigure[Posterior Probability of Infection]{\includegraphics[width=.4\linewidth]{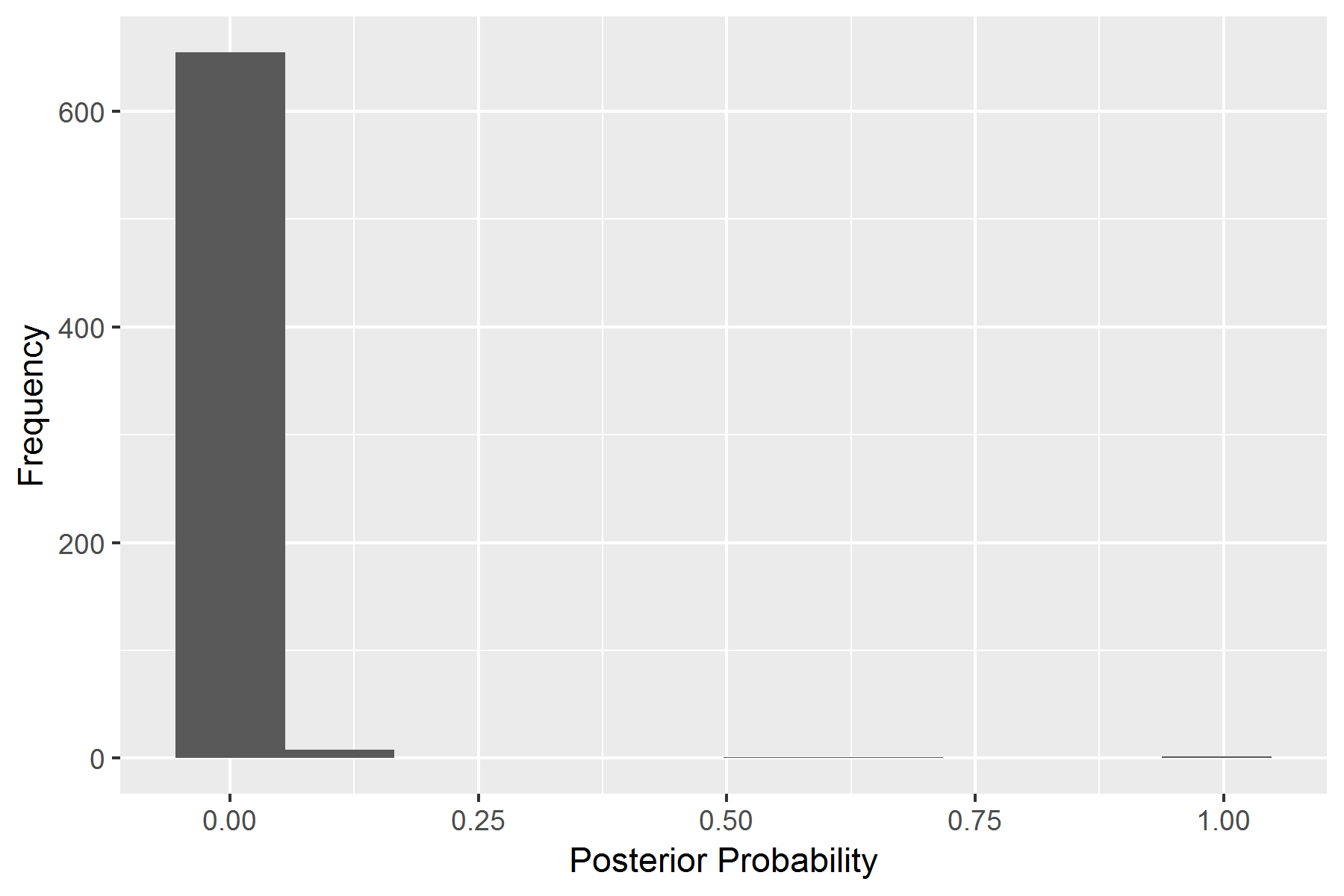}}
    \subfigure[Posterior Probability of Infection >1\%]{\includegraphics[width=.4\linewidth]{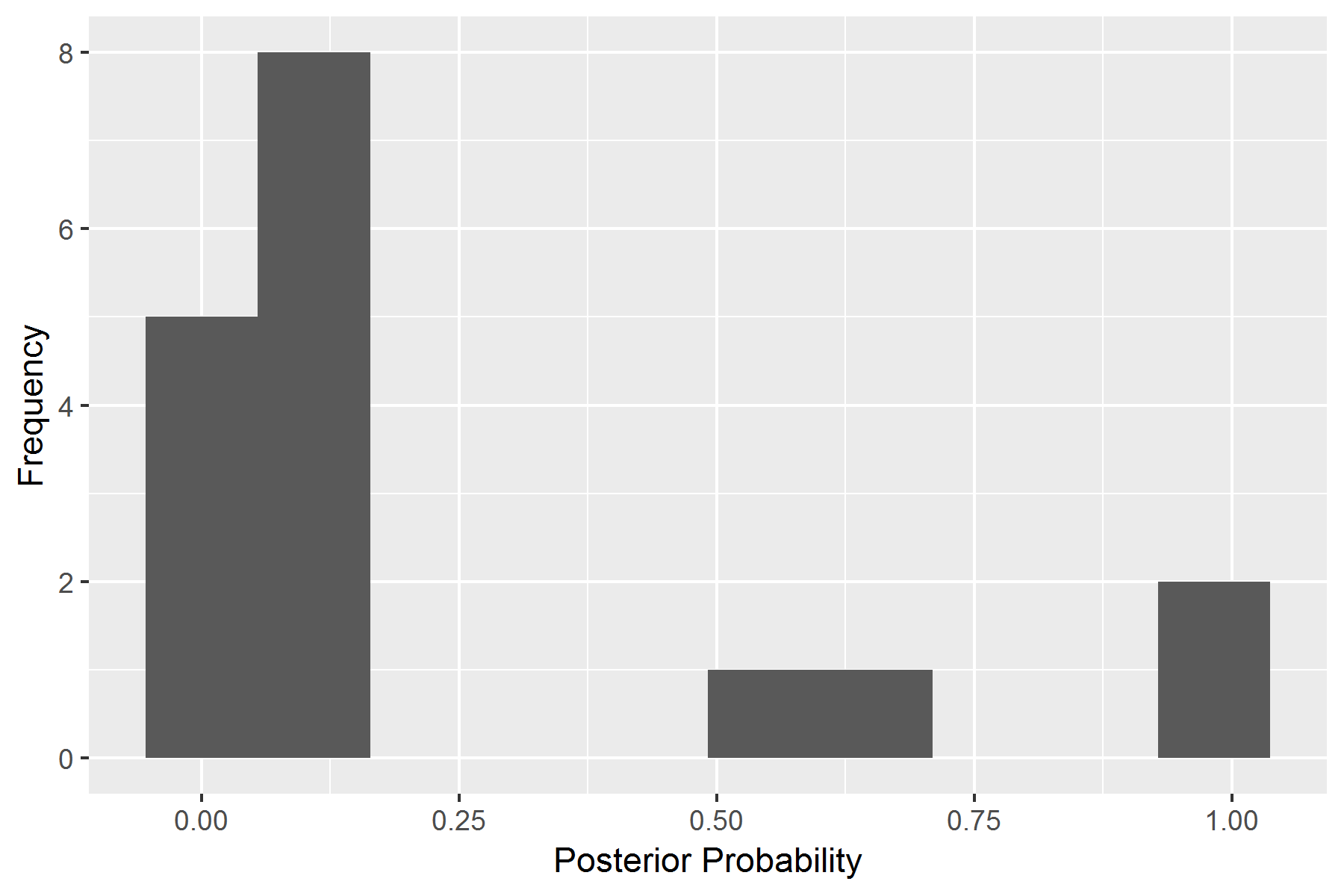}}
    \caption{Posterior probability of past infection of CV19 for all participants and limited to those participants with a probability of greater than 1\%.}
    \label{fig:probs}
\end{figure}

Our primary goal is to estimate the statewide prevalence of prior infection with CV19. Based on our model, the posterior mean prevalence is 1.3\% with a 95\% credible interval of (0.2\%, 2.7\%).  This corresponds to approximately 117,000 adults.  As noted in Section \ref{sec:methods}, the prevalence estimates are based on the latent infection status estimates inferred from the antibody test results.  In Figure \ref{fig:probs}, we show the posterior mean probability of prior infection for all 667 participants included in the analysis.  This illustrates that there were very few participants in the study that were estimated to actually have prior infection. In fact, only 17 (2.5\%) participants had a posterior probability of past infection greater than 1\%.  As expected, those with the highest estimated probabilities were those for whom there was agreement across the tests.  In general, there was little observed agreement across the 3 diagnostic tests.  Of the 39 participants with at least one positive result, only 3 (7.7\%) had a positive result on more than one test. For parameters that had informative prior distributions, Figures S1 and S2 show prior and posterior densities to illustrate learning based on the observed data. 

\begin{figure}
\centering
\includegraphics[width=.8\linewidth]{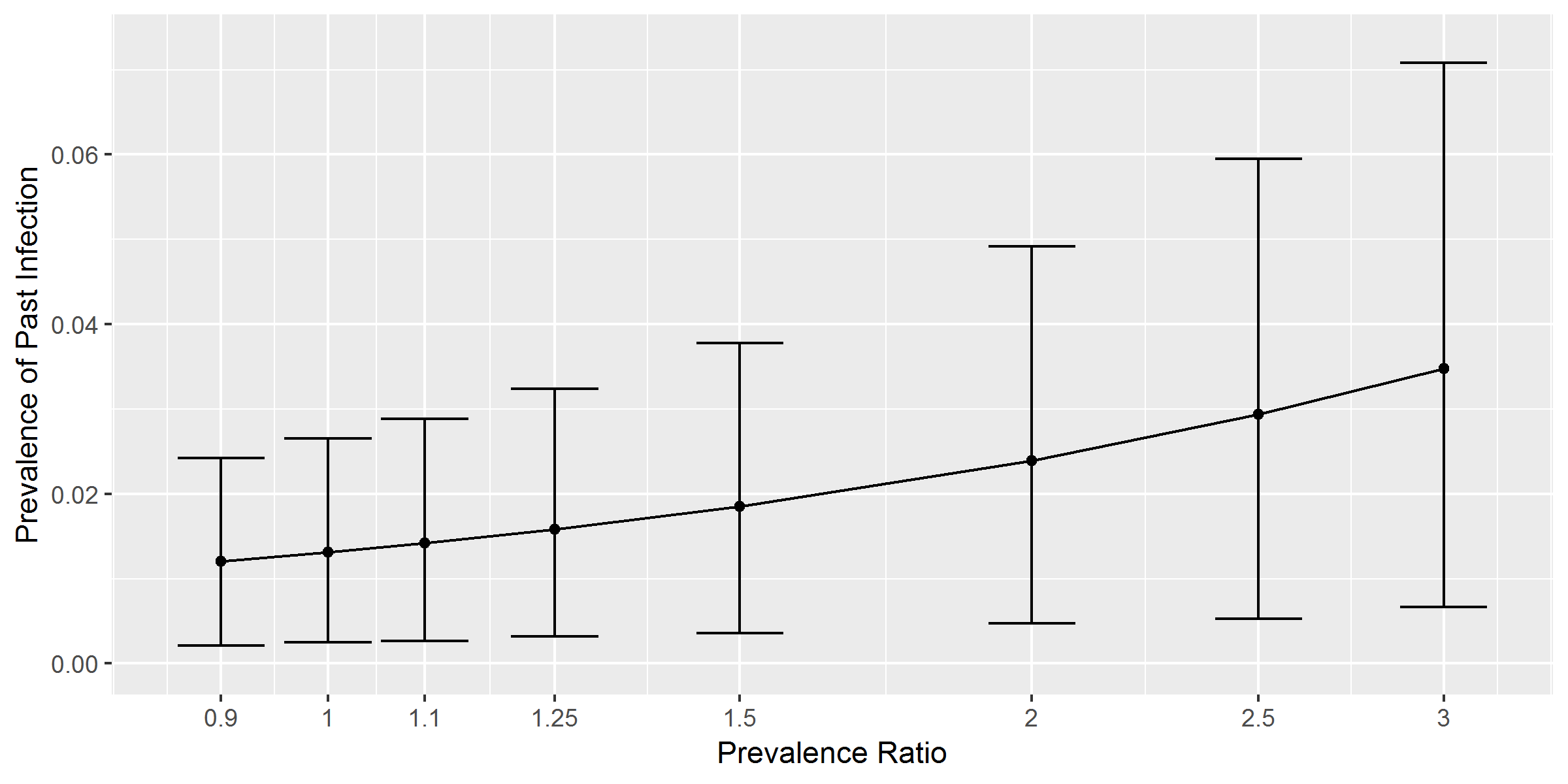}
\caption{Posterior mean and 95\% highest posterior density credible intervals for the prevalence of past infection under several scenarios of non-ignorable non-response. The prevalence ratio is $\lambda$ in Equation [\ref{eq:non}].}
\label{fig:estimates}
\end{figure}

We now turn to the sensitivity analysis that accounted for potential non-ignorable non-response.  The household-level non-response rates by region are shown in Table \ref{tab:response}.  We believe that non-responders were most likely to have a higher prevalence of past infection than responders.  In Figure \ref{fig:estimates}, we show that if we assume the prevalence of past infection in non-responders was three times that of responders, the upper bound of the 95\% credible interval is at a prevalence of 7\%.  Thus, across the range of reasonable scenarios that we considered, we observe rates of past infection that do not dramatically differ from the estimates in our primary analysis.  In addition, all of the estimates in the sensitivity analysis would be aligned with a similar public health and policy response as they are not compatible with having reached herd immunity.

\begin{table}
    \centering
    \caption{Household-level non-response rates by region.}
    \begin{tabular}{lrr}
        Region & Rate \\
        \hline
        Central & 0.783 \\
        East Central & 0.841 \\
        Northeast & 0.840 \\
        Northwest & 0.832 \\
        Southeast & 0.752 \\
        Southeast Central & 0.790 \\
        Southwest & 0.810 \\
        West Central & 0.818 \\
        \hline
    \end{tabular}
    \label{tab:response}
\end{table}

\section{Discussion}
\label{sec:discuss}

In this paper, we present an approach to coherently integrate multiple imperfect diagnostic tests and a model-based analysis.  By using this approach, we are able to estimate the past prevalence of CV19 in the State of Ohio in the United States while appropriately accounting for uncertainty in multiple antibody tests and leveraging the strengths of a designed survey.  Through the Bayesian paradigm, we are also able to incorporate external information through informative prior distributions and provide exact inferences for the prevalence estimates.  Our approach provides policy makers with the best possible estimates of prior infection given the survey results and our current knowledge of the quality of the antibody tests.

Through our study, we estimate the magnitude of prior CV19 infection to be low, roughly 117,000 adults.  However, this still reflects a higher burden of infection than the reported number of cases -- approximately 95,300 after removing deaths through the end of July 2020.  Our sensitivity analysis shows that even with a prevalence ratio of 3 between non-responders and responders, the upper bound of the credible interval is a prevalence of about 7\%.  Thus, even in hypothetical scenarios with large selective response, we conclude that the vast majority of Ohio adults still remain susceptible to CV19 infection.  This implies that the state must remain vigilant and continue to deploy non-pharmaceutical interventions like masks and social distancing to limit the spread of CV19.

Methodologically, we developed a coherent statistical framework for analyzing seroprevalence surveys with multiple imperfect diagnostic tests. Typically, one might rely on a single test or the creation of deterministic rules that combine test results to assess positive cases.  In contrast, we utilized existing methodology for combining the results of multiple diagnostic tests to fully incorporate information on infection status from each test. We then connected the estimated infection status to a multilevel regression and poststratification approach for generating population-based estimates. Through our fully Bayesian, model-based analysis, we are able to appropriately propagate uncertainty and provide exact inferences while synthesizing the information from each type of test administered.  Our approach allows us to make full use of the data collected by the survey while also accounting for its quality.

Our study has significant strengths: a representative random sample; employing multiple diagnostic tests; and conducting a fully Bayesian analysis. There are also limitations. The survey was designed to estimate CV19 prevalence at the level of the whole state and was not designed for estimation within any subgroups or smaller geographies.  A future study should consider oversampling high risk subgroups or regions to enable inference specific to those important populations.  Non-response was a major issue that was addressed in the field using all available resources. We do not have detailed information on non-responders that could have been incorporated into a more sophisticated model for missing data. However, based on our sensitivity analysis, we do not believe the substantive policy implications of the study would change based on a more complicated missing data model. Finally, the informative prior distributions for antibody test characteristics are based on validation data presented in the test inserts -- we are unable to verify the rigor and quality of those validation studies.

In conclusion, we have developed a new statistical analysis framework for analyzing seroprevalence survey data with few positive cases and results from multiple imperfect diagnostic tests. Since estimates of prior and current prevalence are relevant to policy, seroprevalence studies of CV19 are becoming increasingly important and more frequently conducted.  Our methodological approach is a critical component to ensuring that serology data are analyzed in a way that is consistent both with the design of the survey and with the inherent limitations in the accuracy of antibody tests.  This enables policy makers to have access to the best available estimates that also fully and honestly account for all of the sources of uncertainty that contribute to the quantification of CV19 infection.


\end{document}


\maketitle

\begin{figure}
\centering
\subfigure[Sensitivity of Abbott IgG]{
    \includegraphics[width=0.4\textwidth]{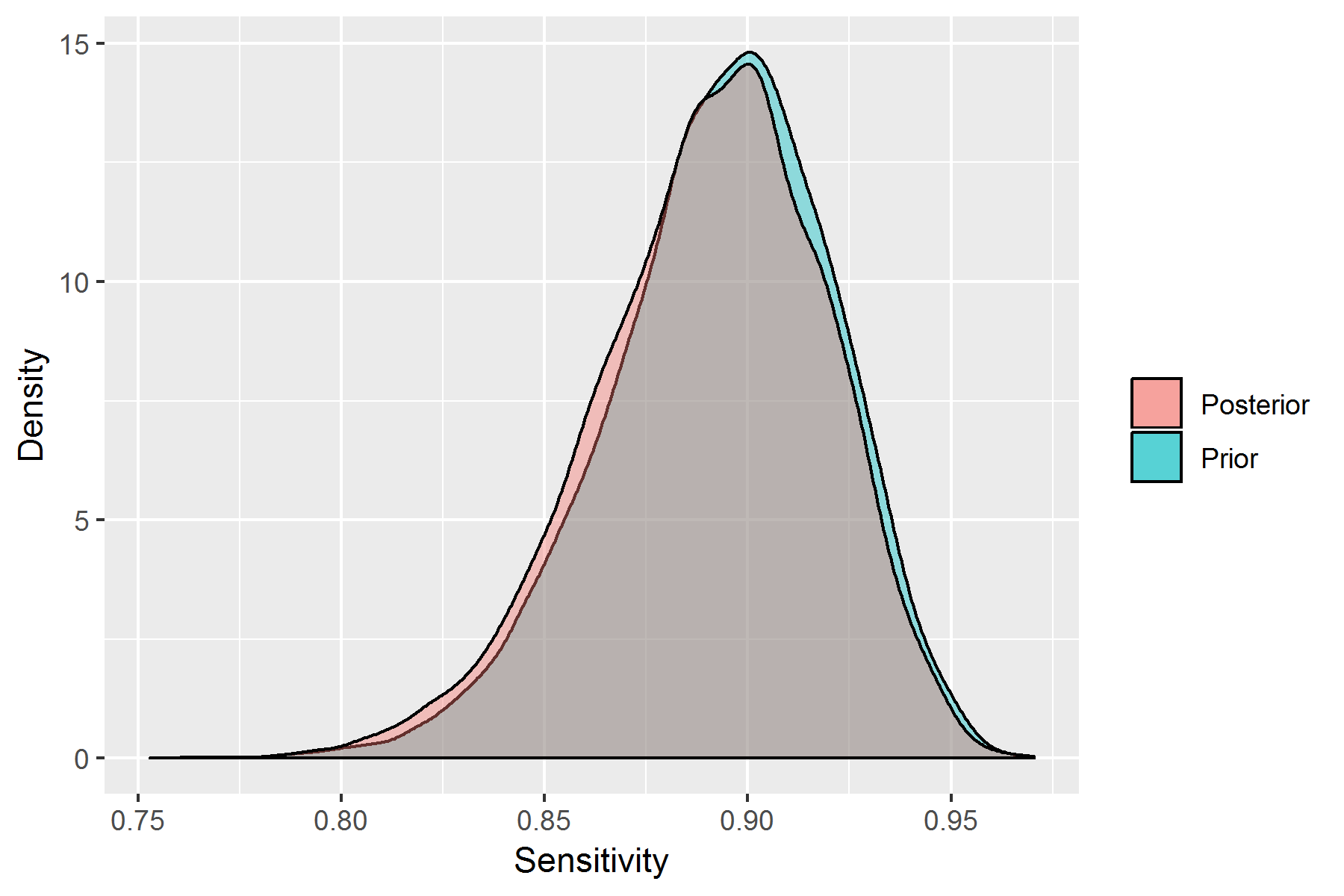}
}
\subfigure[Specificity of Abbott IgG]{
    \includegraphics[width=0.4\textwidth]{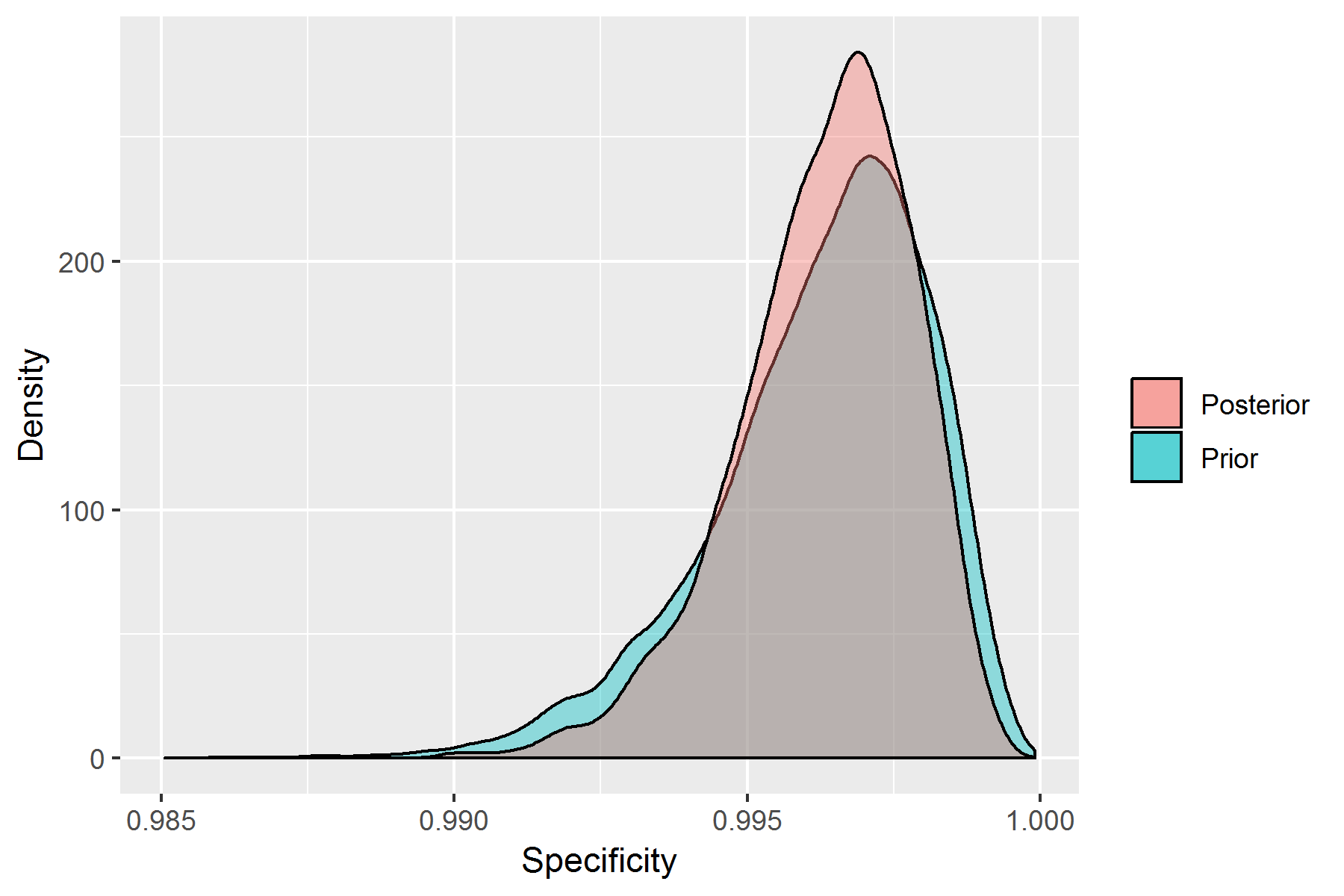}
}
\subfigure[Sensitivity of Liaison IgG]{
    \includegraphics[width=0.4\textwidth]{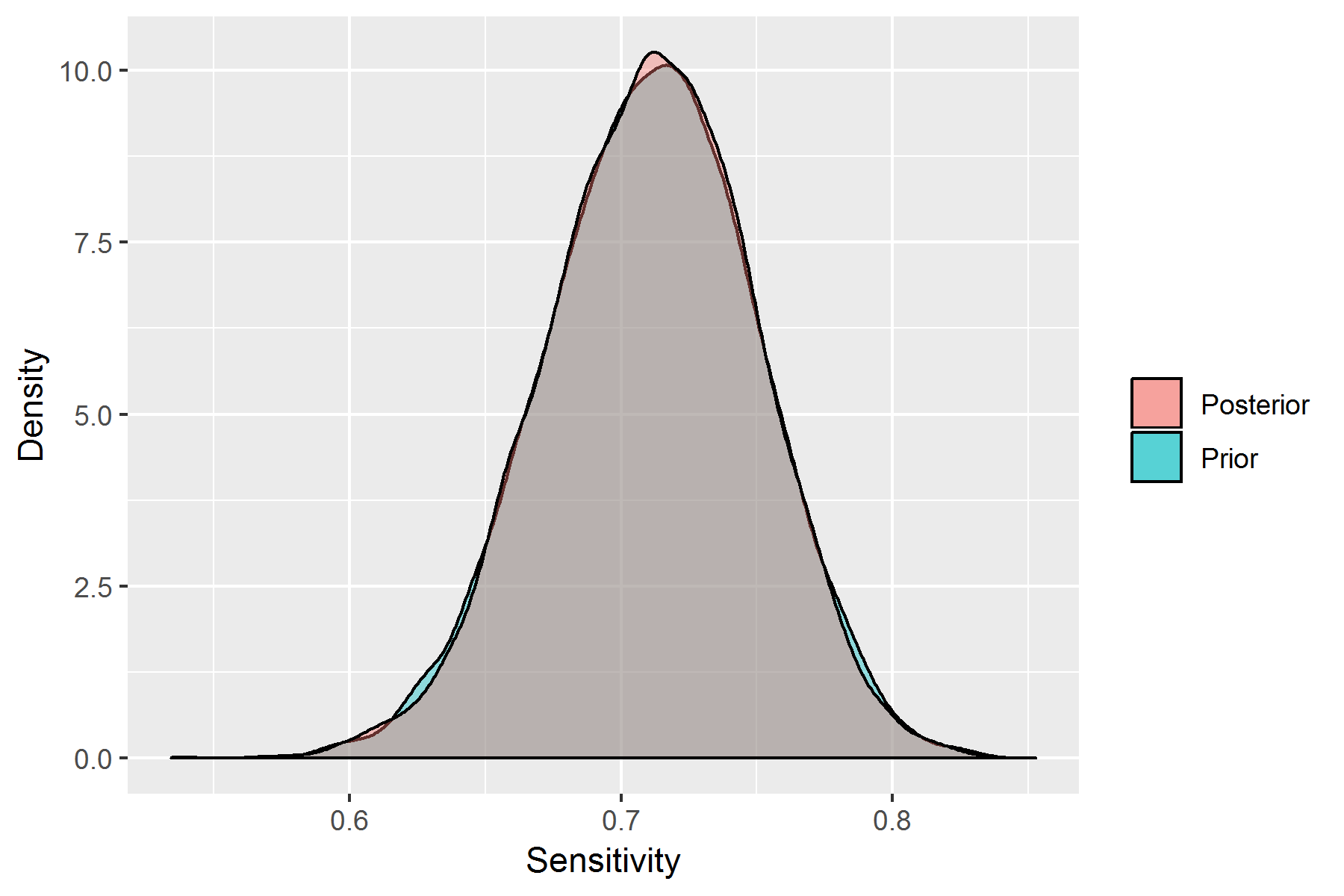}
}
\subfigure[Specificity of Liaison IgG]{
    \includegraphics[width=0.4\textwidth]{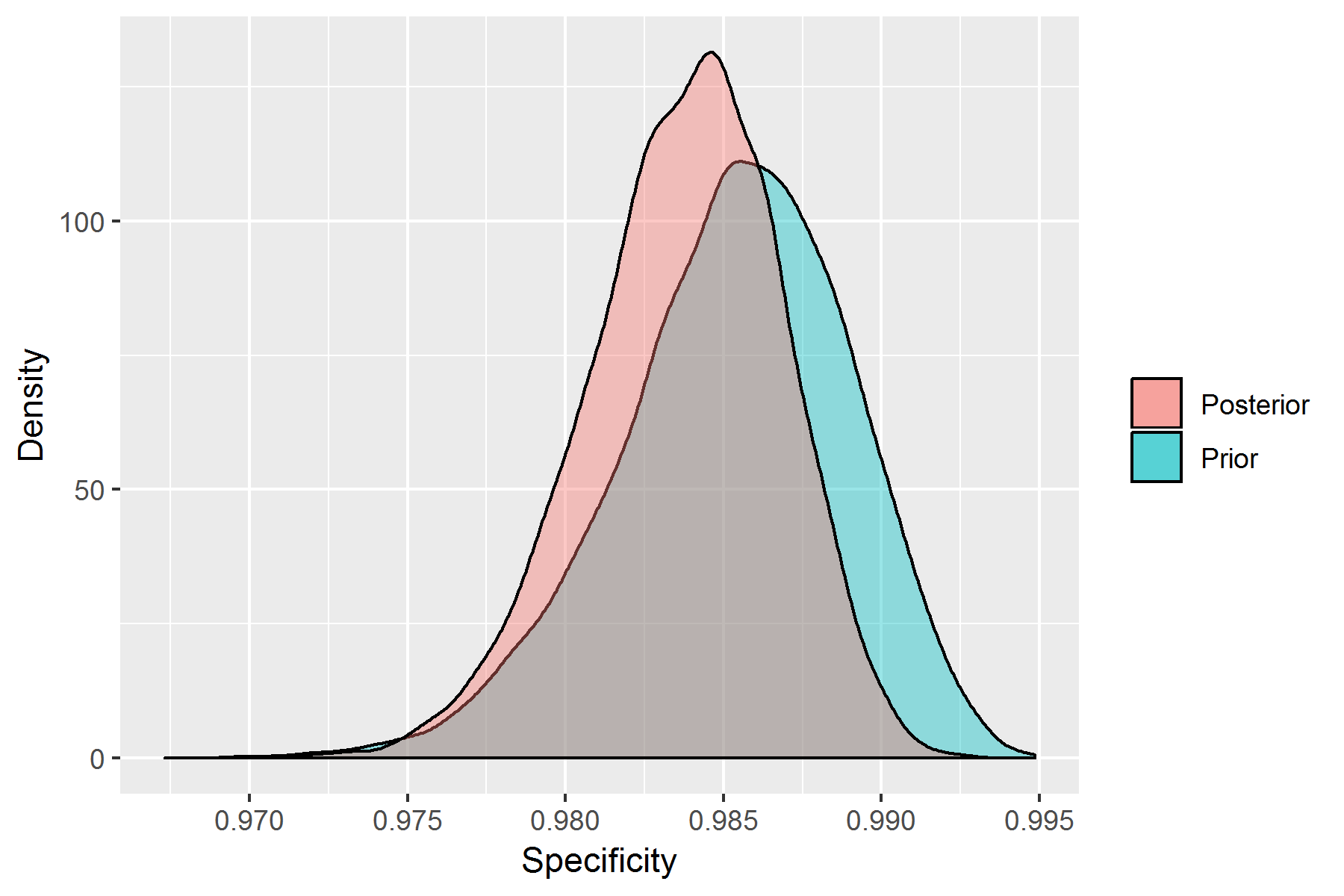}
}
\subfigure[Sensitivity of Epitope IgM]{
    \includegraphics[width=0.4\textwidth]{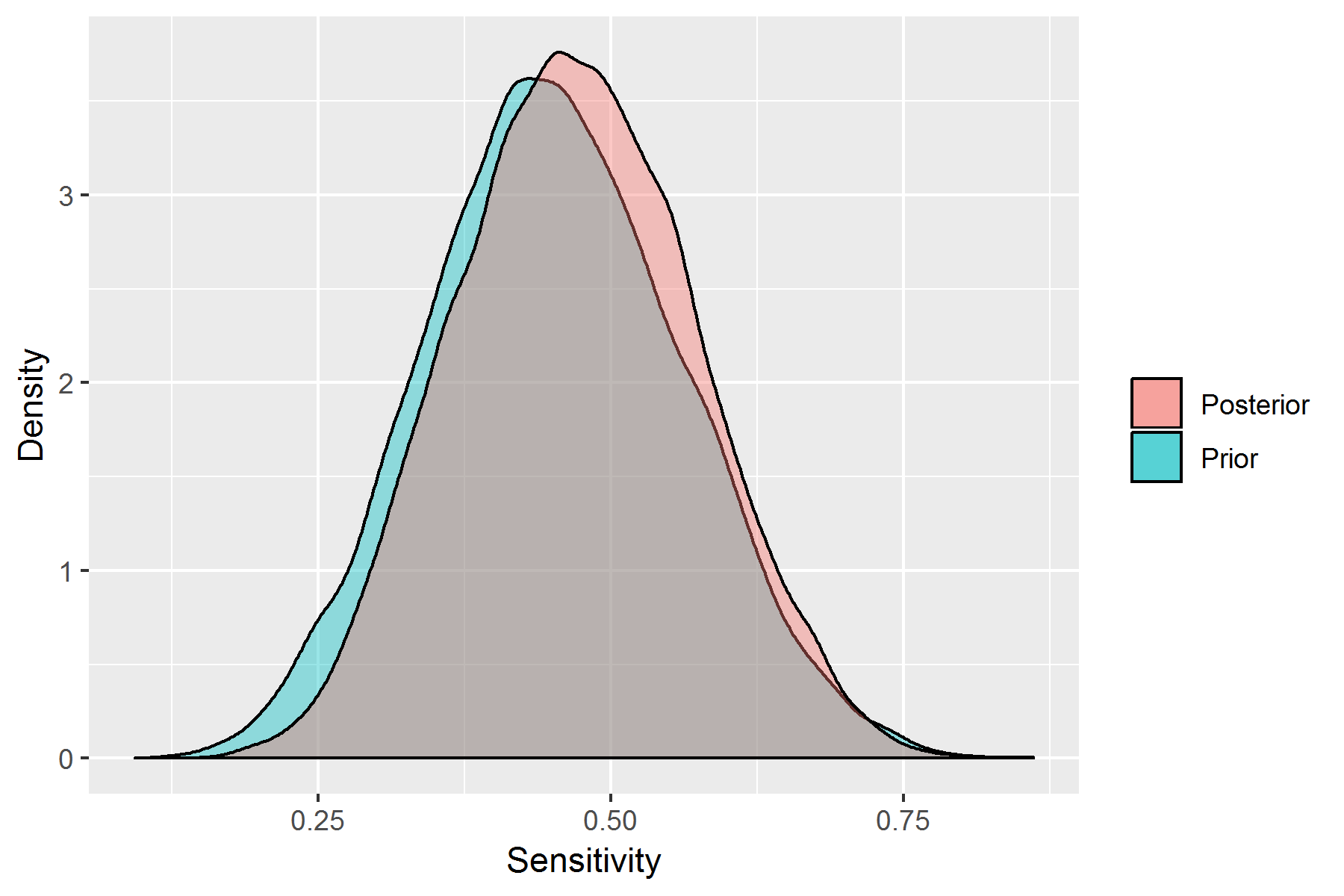}
}
\subfigure[Specificity of Epitope IgM]{
    \includegraphics[width=0.4\textwidth]{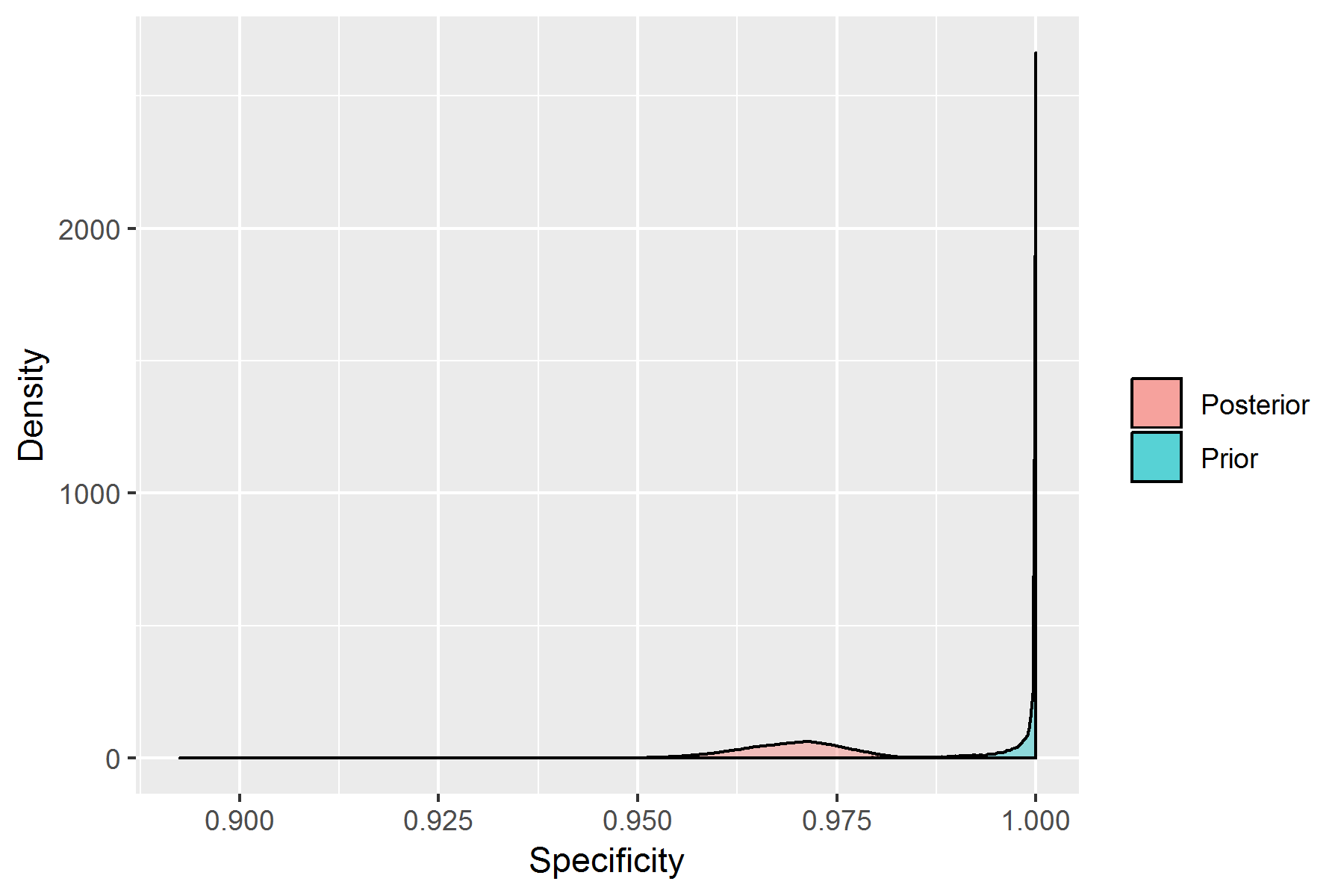}
}
\caption{Density plots of the prior and posterior distributions for the sensitivity and specificity of each antibody test}
\end{figure}

\begin{figure}
\centering
\includegraphics{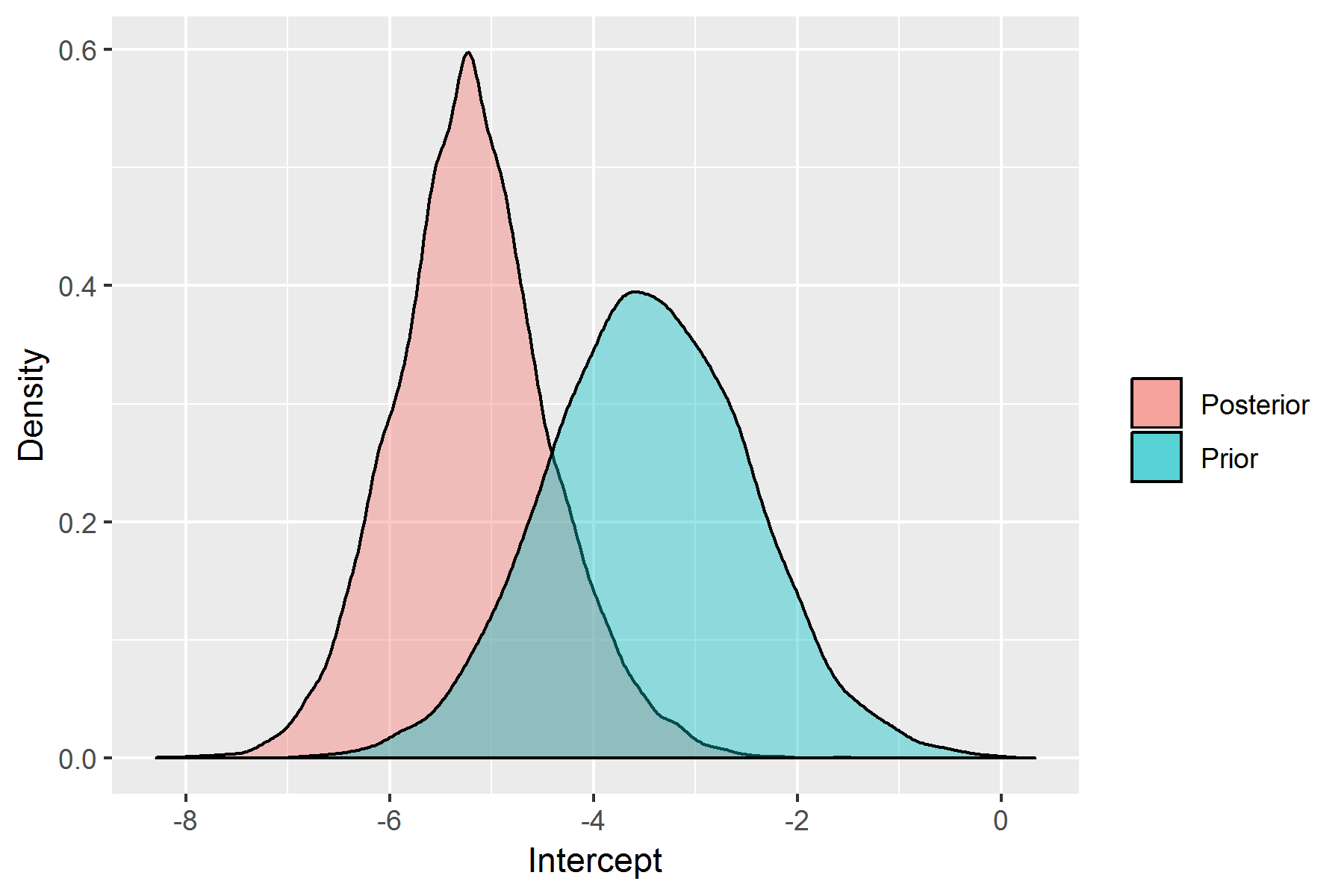}
\caption{Density plots of the prior and posterior distributions of $\alpha$}
\end{figure}